\documentstyle[12pt,epsfig]{article}
\begin{document}

\def\gsim{{~\raise.15em\hbox{$>$}\kern-.85em
          \lower.35em\hbox{$\sim$}~}}
\def\lsim{{~\raise.15em\hbox{$<$}\kern-.85em
          \lower.35em\hbox{$\sim$}~}}
\begin{titlepage}
\title{\Large{Phenomenological Constraints on Supersymmetric Models\\
 with an Anomalous $U(1)$ Flavor Symmetry}}
\author{Galit Eyal \\
\small\it{Department of Particle Physics, Weizmann Institute of
 Science, Rehovot 76100, Israel}}
\date{\small{WIS-99/10/MAR-DPP}}
\maketitle

We investigate supersymmetric models in which an anomalous $U(1)_{X}$
symmetry explains the Yukawa hierarchy, and the related $D_{X}$-term
plays a role in supersymmetry breaking. We use a bottom-up approach to 
model building. Phenomenological viability leads to a scenario with
degenerate squark and slepton spectra and with heavy gauginos. Features of
a K\"ahler potential that allows for such a scenario are described. \\
\vspace{0.5cm}
\end{titlepage}
 
\section{Introduction}

Two open questions in the framework of Supersymmetry (SUSY) are the
mechanism of SUSY breaking and the structure of flavor parameters. In a
particularly interesting class of models, a single, anomalous $U(1)_{X}$
flavor symmetry plays an important role in answering these two questions.

SUSY breaking with an anomalous $U(1)_{X}$ has been introduced in 
refs.~\cite{a:dp96,a:bd96}, and further analyzed in ref.~\cite{a:adm98}.
This mechanism includes a $U(1)_{X}$ gauge symmetry which is anomalous due
to a non-vanishing value of
\begin{equation}
\delta_{GS}=\frac{1}{192\pi^{2}}\sum_{i}q_{X_{i}}, \label{eq:delgs}
\end{equation} 
where $q_{X_{i}}$ is the $U(1)_{X}$ charge of the $i$ chiral superfield. 
Without loss of generality $\delta_{GS}$ is taken to be positive. The 
anomaly cancels by a non-trivial transformation of the dilaton ($S$)
superfield according to the Green-Schwarz mechanism~\cite{a:gs84}. This
results in the generation of a Fayet-Iliopoulos term:
\begin{equation}
\xi^{2}=-\frac{1}{2}\delta_{GS}M_{p}^{2}K', \label{eq:K'}
\end{equation}
where $K$ is the dimensionless K\"ahler potential and $K'\equiv
\frac{\partial K}{\partial S}$. All the chiral superfields charged under 
the Standard Model (SM) gauge group are assigned positive (or zero) 
charges under $U(1)_{X}$, in order not to break any of these symmetries at
a high-energy scale. There is, however, a single SM-singlet field with a 
negative $U(1)_{X}$ charge $(\phi_{-})$. Minimization of the scalar
potential of the complete model results in SUSY breaking with the
following non-zero vacuum expectation values (vevs):
\begin{equation}
\left< \phi_{-}\right>,\hspace{1cm} \left< F_{S}\right> ,\hspace{1cm}
\left< D_{X}\right>. \label{eq:vevs}
\end{equation}
Depending on the details of the model, there might be additional fields
that receive non-zero vevs.

The $U(1)_{X}$ symmetry breaking produces a naturally small parameter:
\begin{equation}
\epsilon=\frac{\left< \phi_{-}\right>}{M_{p}}\sim\frac{\xi}{M_{p}}.
\end{equation}
This small parameter motivates the use of the $U(1)_{X}$ as a horizontal
flavor 
symmetry~\cite{a:ir94}-\cite{a:nw97}. We take $\epsilon$ to be of
order of the Cabbibo angle, $\lambda \sim 0.2$. (In our framework
$\delta_{GS}= O(\lambda^{2}-\lambda)$, which implies $K'= O(1-\lambda)$.)
The anomalous $U(1)_{X}$ is an interesting flavor symmetry since the
$D_{X}$-term contributes to the soft masses of light scalar fields which
carry a $U(1)_{X}$ charge and induces non-degeneracy among them.

In this work we use a bottom-up approach to analyze models with a single
anomalous $U(1)_{X}$ that acts as a horizontal symmetry. (We assume that 
there are no additional non-anomalous horizontal $U(1)$s.) The requirement
that the $U(1)_{X}$ symmetry explains the mass parameters in the fermion
sector leads to an almost unique charge assignment for the 
Supersymmetric-SM matter fields. This charge assignment leads to only mild 
alignment between the fermions and sfermions~\cite{a:ns93}. We assume that
all sfermion masses are below, or just at, the TeV scale. (Models with
heavy first two sfermion generations, the motivation for them and their
potential problems, have been discussed in the 
literature~\cite{a:nw97},\cite{a:dg95}-\cite{a:hkn98}. 
We do not consider
them here.) Consequently, in order to satisfy phenomenological 
constraints, the squark spectrum as well as the slepton spectrum have to 
be roughly degenerate at low-energy, so the horizontal symmetry is not
manifest in the squark and slepton spectra. We describe the possible
high-energy scenarios that produce such a spectrum. Assuming that the
high-energy parameters are determined by the vevs in eq.~(\ref{eq:vevs}),
the required relations between the $F_{S}$-term, $D_{X}$-term and the
derivatives of the K\"ahler potential that need to be satisfied, within
the different scenarios, are analyzed. 

In section~\ref{s:frame} we describe the general framework in which we
work. In section~\ref{s:fermion} the choice of charge assignments is
explained. Two specific examples and a general analysis of the
phenomenologically viable scenarios are given in section~\ref{s:sfermion}.
Section~\ref{s:conclu} contains some concluding remarks.

\section{The Framework \label{s:frame}}

We assume that the leading contributions to the relevant dimensionful 
parameters come from the vevs in eq.~(\ref{eq:vevs}). Any contributions by 
additional vevs for scalar fields or $F$-terms are sub-dominant. Then we
can make the following statements concerning the flavor and Higgs
parameters at 
high-energy~\cite{a:sw83}-\cite{a:kl93}:

(i) The Yukawa parameters depend on the $U(1)_{X}$ charges of the matter
fields~\cite{a:lns94}. For example, consider the following down quark mass
matrix element:
\begin{equation}
Y^{d}_{11}\phi_{d}q_{1}\bar{d}_{1}=
\frac{a^{d}_{11}}{M_{p}^{n}}\phi_{-}^{n}\phi_{d}q_{1}\bar{d}_{1}
\rightarrow
M_{11}^{d}=a^{d}_{11}\epsilon^{n}\left< \phi_{d}\right>.
\end{equation}
We use $q_{i}$ to denote the quark doublets, $\bar{d}_{i}$ ($\bar{u}_{i}$)
the down (up) quark singlets, $\ell_{i}$ the lepton doublets, 
$\bar{e}_{i}$ the charged lepton singlets, and $\phi_{u}$ and $\phi_{d}$
the Higgs doublet fields. $Y^{f}_{ij}$ denote Yukawa couplings, 
$a^{f}_{ij}$ are $O(1)$ coefficients, and $n=q_{X\phi d}+q_{Xq_{1}}
+q_{X\bar{d}_{1}}$.

(ii) Diagonal elements of the sfermion mass-squared matrices,
$\tilde{m}_{i}^{o\:2}$, receive $\tilde{m}_{3/2}$ (flavor independent,
universal) and $D_{X}$ (flavor dependent) contributions as follows:
\begin{equation}
\tilde{m}^{o\:{2}}_{i}=\tilde{m}^{2}_{3/2}-q_{X_{i}}\left<
D_{X}\right> 
\label{eq:msq}.
\end{equation}
When the vanishing of the cosmological constant is imposed, the gravitino
mass $\tilde{m}_{3/2}$ is given by: 
\begin{equation}
\tilde{m}^{2}_{3/2}=\frac{1}{3}K''|\left< F_{S} \right> |^{2}.
\label{eq:m32}
\end{equation}

(iii) Gaugino masses, $\tilde{m}^{o}_{1/2}$, receive universal
contributions:
\begin{equation}
\tilde{m}^{o}_{1/2}=\frac{\left<F_{S}\right>}{\left<S+S^{*}\right>}.
\label{eq:mo12}
\end{equation}
Below we take $\left< S\right>\simeq 2$.

(iv) Off-diagonal elements in the mass-squared matrices are suppressed
compared to the diagonal terms for two reasons: First, they receive 
contributions proportional to $F$-terms smaller than $F_{S}$, and, second,
they are suppressed by the horizontal symmetry.

(v) A-terms are proportional to the Yukawa couplings~\cite{a:kl93}:
\begin{equation}
A^{f}_{ij}=\left< F_{S}\right> K' Y^{f}_{ij}.
\end{equation}
Using eq.~(\ref{eq:mo12}) we get:
\begin{equation}
A^{f}_{ij}= \left<S+S^{*}\right>\tilde{m}^{o}_{1/2}K' Y^{f}_{ij}.
\end{equation}
Later we take $Y^{u}_{33}=O(1)$. In order to avoid large contributions
by the corresponding A-term to the renormalization group equations (RGE) 
that might lead to a negative stop mass, we need to impose (especially
when the gauginos are heavy) $K'\leq O(\lambda)$. Once this is imposed,
the A-term contributions are negligible.

(vi) Higgs sector mass parameters:\\
We assume that the $\mu$ term is not generated in the superpotential. It
can be generated by the Giudice-Masiero mechanism~\cite{a:gm88,a:n95}:
\begin{equation}
\mu=a_{\mu}\epsilon^{q_{Xh_{d}}}\tilde{m}_{3/2},
\end{equation}
where $a_{\mu}$ is an $O(1)$ coefficient and the suppression by powers of
$\epsilon$ comes from the need to make the term $U(1)_{X}$ symmetric. We
always set $q_{Xq_{3}}=q_{X\bar{u}_{3}}=q_{Xh_{u}}=0$ in order to have 
$Y^{u}_{33}= O(1)$, and $q_{Xh_{d}} \geq 0 $ in order to avoid breaking
the SM symmetry at high-energy. The other relevant Higgs parameters 
are given by~\cite{a:kl93}
\begin{eqnarray}
B&=&2\mu \tilde{m}_{3/2},\\
m_{h_{u}}^{o\:2}&=&\tilde{m}^{2}_{3/2},\\
m_{h_{d}}^{o\:2}&=&\tilde{m}^{2}_{3/2} -q_{Xh_{d}}\left< D_{X}\right>,\\
\tan\beta&\sim&\frac{m^{o\:2}_{h_{u}}+m^{o\:2}_{h_{d}}}{B}.
\end{eqnarray}
If the Higgs masses are not dominated by the $D_{X}$-term then
$\tan\beta =O(\epsilon^{-q_{Xh_{d}}})$.

The effects of running from high-energy to low-energy are implemented
following ref.~\cite{a:ceklp95}. We define an average high-energy squark
mass-squared $\tilde{m}^{o\:{2}}_{q}$ and an average high-energy slepton
mass-squared $\tilde{m}^{o\:{2}}_{\ell}$. We denote:
\begin{equation}
X^{o}_{q}=\frac{\tilde{m}^{o\:{2}}_{1/2}}{\tilde{m}_{q}^{o\:{2}}},
\hspace{2cm} X^{o}_{\ell}=
\frac{\tilde{m}^{o\:{2}}_{1/2}}{\tilde{m}_{\ell}^{o\:{2}}}. \label{eq:Xo}
\end{equation}
At the low-energy scale we get the following average values (neglecting
contributions from A-terms and $O(m^{2}_{Z})$ corrections):
\begin{eqnarray}
\tilde{m}^{2}_{q}&\simeq
&\tilde{m}^{o\:{2}}_{q}+7\tilde{m}^{o\:{2}}_{1/2}, \label{eq:runsq} \\
\tilde{m}^{2}_{\ell}&\simeq&\tilde{m}^{o\:{2}}_{\ell}+
0.3\tilde{m}^{o\:{2}}_{1/2},\label{eq:runsl}\\
X_{q}&=&\frac{\tilde{m}_{3}^{2}}{\tilde{m}_{q}^{2}}\simeq
\frac{9X^{o}_{q}}{1+7X^{o}_{q}} \rightarrow X_{q}=[0,\frac{9}{7}),\\
X_{\ell}&=&\frac{\tilde{m}_{1}^{2}}{\tilde{m}_{\ell}^{2}}\simeq
\frac{0.16X^{o}_{\ell}}{1+0.3X^{o}_{\ell}} \rightarrow
X_{\ell}=[0,0.53).
\end{eqnarray}

The quark and lepton $U(1)_{X}$ charges are assigned in such a way that 
they reproduce the flavor parameters:   
\begin{eqnarray}
(m_{u}:m_{c}:m_{t})&\sim & (\lambda^{7}:\lambda^{4}:1),\\   
(m_{d}:m_{s}:m_{b})&\sim & (\lambda^{7}:\lambda^{5}:\lambda^{3}),\\
(m_{e}:m_{\mu}:m_{\tau})&\sim &
(\lambda^{8}-\lambda^{9}:\lambda^{5}:\lambda^{3}),\\
|V_{us}|\sim \lambda, \hspace{0.5cm}& 
|V_{cb}|\sim \lambda^{2},& \hspace{0.5cm} 
|V_{ub}|\sim \lambda^{3}. 
\end{eqnarray}
We do not impose additional constraints on the charges, as required by the
Green-Schwarz mechanism for example, since we assume that there could be
additional superheavy matter fields in the model that are vector-like 
under the SM gauge group but chiral under $U(1)_{X}$. $\tan\beta$ is in
the range $1-\lambda^{-2}$. We found no reason to prefer one choice over
the other.

\section{The Fermion Sector \label{s:fermion}}

In our examples we choose $q_{Xh_{d}}= 0$ and consequently $\tan\beta \sim
1$. With our requirements the charges of the quarks are fixed uniquely, 
but there is still freedom left in the lepton sector. Different lepton
charges lead to different contributions to Flavor Changing Neutral
Currents (FCNC). The charges we choose to use are given in
table~\ref{t:charges}. 
\begin{table}[h] 
\begin{center}
\begin{tabular}{|ccc|ccc|ccc|ccc|ccc|}
\hline
$q_{1}$ & $q_{2}$ & $q_{3}$ & $\bar{u}_{1}$& $\bar{u}_{2}$&$\bar{u}_{3}$
& $\bar{d}_{1}$& $\bar{d}_{2}$ & $\bar{d}_{3}$ &  $\ell_{1}$ & 
$\ell_{2}$ & $\ell_{3}$ & 
$\bar{e}_{1}$& $\bar{e}_{2}$& $\bar{e}_{3}$\\
\hline
$3$& $2$& $0$ & $4$& $2$& $0$ &$4$& $3$& $3$ & $5$& $3$& $2$ & $4$ &
$2$& $1$ \\
\hline
\end{tabular}
\end{center}
\caption{$U(1)_{x}$ charges of the superfields.}
\label{t:charges}
\end{table}
The choice $q_{X\ell_{1}}=4$, $q_{X\ell_{2}}=q_{X\ell_{3}}=3$ and
$q_{X\bar{e}_{1}}=5$ would have led, for example, to larger contributions
to $Br(\mu\rightarrow e\gamma)$. A different charge assignment was given
in ref.~\cite{a:nw97}, but there a deviation of factors up to $O(10)$ in
the mass ratios was allowed, and a spectrum of heavy squarks was produced.

With our charge assignment the following mass matrices are produced:
\begin{equation}
M^{u}\sim \left< \phi_{u}\right> \left( \begin{array}{ccc}
\epsilon^{7} & \epsilon^{5} & \epsilon^{3} \\
\epsilon^{6} & \epsilon^{4} & \epsilon^{2} \\
\epsilon^{4} & \epsilon^{2} & 1
\end{array} \right), \hspace{1cm}
M^{d}\sim \left<\phi_{d}\right> \left( \begin{array}{ccc}
\epsilon^{7} & \epsilon^{6} & \epsilon^{6} \\
\epsilon^{6} & \epsilon^{5} & \epsilon^{5} \\
\epsilon^{4} & \epsilon^{3} & \epsilon^{3}
\end{array} \right),
\end{equation}
\begin{equation}
M^{\ell}\sim \left< \phi_{d}\right> \left( \begin{array}{ccc}
\epsilon^{9} & \epsilon^{7} & \epsilon^{6} \\
\epsilon^{7} & \epsilon^{5} & \epsilon^{4} \\
\epsilon^{6} & \epsilon^{4} & \epsilon^{3}
\end{array} \right).
\end{equation}
This is the form of the matrices at high energy. The diagonalizing
matrices for the fermions are:
\begin{equation}
V_{L}^{u}\sim
\left(\begin{array}{ccc}
1 & \epsilon & \epsilon^{3}\\
\epsilon & 1 & \epsilon^{2}\\
\epsilon^{3} & \epsilon^{2} & 1 \end{array} \right),
\hspace{1cm}
V_{R}^{u}\sim
\left(\begin{array}{ccc}
1 & \epsilon^{2} & \epsilon^{4}\\
\epsilon^{2} & 1 & \epsilon^{2}\\
\epsilon^{4} & \epsilon^{2} & 1 \end{array} \right),
\end{equation}
\begin{equation}
V_{L}^{d}\sim
\left(\begin{array}{ccc}
1 & \epsilon & \epsilon^{3}\\
\epsilon & 1 & \epsilon^{2}\\
\epsilon^{3} & \epsilon^{2} & 1 \end{array} \right),
\hspace{1cm}
V_{R}^{d}\sim
\left(\begin{array}{ccc}
1 & \epsilon & \epsilon\\
\epsilon & 1 & 1\\
\epsilon & 1 & 1 \end{array} \right),
\end{equation}
\begin{equation}
V_{L}^{\ell}\sim V_{R}^{\ell}\sim
\left(\begin{array}{ccc}
1 & \epsilon^{2} & \epsilon^{3}\\
\epsilon^{2} & 1 & \epsilon\\
\epsilon^{3} & \epsilon & 1 \end{array} \right).
\end{equation}
The size of the gaugino-fermion-sfermion flavor changing couplings is 
determined by the above matrices and by the diagonalizing matrices for the
sfermions. However, the off-diagonal entries in the diagonalizing matrices
for the fermions given above are dominant and they determine the 
couplings. Only a mild alignment is found~\cite{a:ns93}.

\section{The Sfermion Sector \label{s:sfermion}}

The charge assignments above lead to non-universal contributions to the
masses of the sfermions at high-energy (eq.~(\ref{eq:msq})). Since in our
scenario there is only mild alignment, and we do not consider the heavy
squark scenario, only a degeneracy among the squarks and among the
sleptons at low energies can avoid too large FCNC. The limits on FCNC
parameters are taken from ref.~\cite{a:ggms96}, updated with the new bound
on $Br(\mu\rightarrow e\gamma)$~\cite{a:m99}, where the $\delta$'s are 
defined as, for example,
\begin{equation}
(\delta^{d}_{RR})_{12}\sim(K^{d}_{R})_{11}(K^{d}_{R})^{*}_{12}
\frac{\tilde{m}^{2}_{d_{R_{1}}}-\tilde{m}^{2}_{d_{R_{2}}}}
{\tilde{m}_{q}^{2}}\sim(V_{R}^{d})^{*}_{12}
\frac{\tilde{m}^{2}_{d_{R_{1}}}-\tilde{m}^{2}_{d_{R_{2}}}}
{\tilde{m}_{q}^{2}}.
\end{equation}
Here $K^{d}_{R}$ denotes gluino couplings to right-handed down quarks and
'right-handed' down squarks. 
 
The assumptions made in section~\ref{s:frame} regarding the size of the
different soft terms, and particularly the A-terms, imply that the limits 
on $(\delta^{f}_{LR})_{ij}$ do not pose additional constraints. 

There are two ways in which to achieve the necessary degeneracy at low
energies:

(i) Heavy gauginos induce degeneracy through RGE (see 
    eqs.~(\ref{eq:runsq})-(\ref{eq:runsl})).

(ii) Large universal contributions to sfermion masses 
     (see eq.~(\ref{eq:msq})).\\
Below we first examine two limiting cases, heavy and light gauginos, and
then give a general analysis of the possible scenarios.

We start by presenting the sfermion mass-squared matrices at high-energy,
allowing for a universal contribution. We give only the diagonal elements
because the off-diagonal ones are, as mentioned above, suppressed: 
\begin{equation}
\tilde{M}^{q^{2}}_{LL}=
\tilde{m}^{2}\left(\begin{array}{ccc}
3+z & & \\
& 2+z & \\ 
& & z 
\end{array}\right), \label{eq:hesq1}
\end{equation}
\begin{equation}
\tilde{M}^{u^{2}}_{RR}= 
\tilde{m}^{2}\left(\begin{array}{ccc}
4+z & & \\
& 2+z & \\ 
& & z 
\end{array}\right), \hspace{2cm}
\tilde{M}^{d^{2}}_{RR}= 
\tilde{m}^{2}\left(\begin{array}{ccc}
4+z & & \\
& 3+z & \\ 
& & 3+z 
\end{array}\right),\label{eq:hesq2}
\end{equation}
\begin{equation}
\tilde{M}^{\ell^{2}}_{LL}= 
\tilde{m}^{2}\left(\begin{array}{ccc}
5+z & & \\
& 3+z & \\ 
& & 2+z 
\end{array}\right), \hspace{2cm}
\tilde{M}^{e^{2}}_{RR}= 
\tilde{m}^{2}\left(\begin{array}{ccc}
4+z & & \\
& 2+z & \\ 
& & 1+z 
\end{array}\right),\label{eq:hesl}
\end{equation}
where 
\begin{equation}
\tilde{m}^{2}=-\left< D_{X}\right>,\hspace{2cm}
z=-\frac{\tilde{m}^{2}_{3/2}}{\left< D_{X}\right>}. \label{eq:m2z}
\end{equation}
Here $\tilde{m}^{o\:{2}}_{q}\simeq(2.2+z)\tilde{m}^{2}$ and
$\tilde{m}^{o\:{2}}_{\ell}\simeq (3+z)\tilde{m}^{2}$.\\ 

\subsection{Heavy Gauginos: $\tilde{m}_{1/2}^{o} \gsim
\tilde{m}^{o}_{q,\ell}$ \label{s:heavy}}

The gauginos are heavy when the following relations are fulfilled
(eqs.~(\ref{eq:msq})-(\ref{eq:mo12})):
\begin{equation}
\frac{\left<F_{S}\right>^{2}}{\left<S+S^{*}\right>^{2}} \gsim
\tilde{m}^{2}_{3/2}\hspace{0.3cm} \rightarrow\hspace{0.3cm} 
\frac{1}{\left<S+S^{*}\right>^{2}} \gsim \frac{1}{3}K'' \hspace{0.3cm} 
\rightarrow \hspace{0.3cm}  K''\leq O(\lambda), 
\end{equation}
and
\begin{equation}
\hspace{0.5cm}\frac{\left<F_{S}\right>^{2}}{\left<S+S^{*}
\right>^{2}}\gsim -\left<D_{X}\right> \hspace{0.3cm} \rightarrow
\hspace{0.3cm} -\frac{\left<D_{X}\right>}
{\left<F_{S}\right>^{2}}<O(\lambda).
\end{equation}
In this scenario the degeneracy of the sfermion masses at low-energy is 
enhanced compared to the degeneracy at high-energy as given by the
mass-squared matrices in eqs.~(\ref{eq:hesq1})-(\ref{eq:hesl}). 

Let us take for example $X^{o}_{q}=4$. The gaugino masses at high-energy 
are estimated to be 
$\tilde{m}_{1/2}^{o\:{2}}=X^{o}_{q}\tilde{m}^{o\:{2}}_{q}$.
The sfermion mass-squared matrices have the following form at low-energy:
\begin{eqnarray}
&diag\{\tilde{M}^{q^{2}}_{LL}\}&\simeq 
\tilde{m}^{2}\{64+29z,\;\;\; 63+29z,\;\;\; 61+29z\}, \label{eq:lehg1} \\
&diag\{\tilde{M}^{u^{2}}_{RR}\}&\simeq 
\tilde{m}^{2}\{65+29z,\;\;\; 63+29z,\;\;\; 61+29z\},\\
&diag\{\tilde{M}^{d^{2}}_{RR}\}&\simeq 
\tilde{m}^{2}\{65+29z,\;\;\; 64+29z,\;\;\; 64+29z\},\\
&diag\{\tilde{M}^{l^{2}}_{LL}\}&\simeq 
\tilde{m}^{2}\{8+2z,\;\;\; 6+2z,\;\;\; 5+2z\},\\
&diag\{\tilde{M}^{e^{2}}_{RR}\}&\simeq 
\tilde{m}^{2}\{7+2z,\;\;\; 5+2z,\;\;\; 4+2z\},\label{eq:lehg5}
\end{eqnarray}
and
\begin{equation}
X_{q}\simeq 1.2, \hspace{2cm} X_{\ell}\simeq 0.3.
\end{equation}
In eqs.~(\ref{eq:lehg1})-(\ref{eq:lehg5}) we give the same correction to
all squarks (eq.~(\ref{eq:runsq})) and to all sleptons
(eq.~(\ref{eq:runsl})). This is an approximation. 'Left-handed', 
'right-handed', up and down sfermions run differently from high-energy to
low-energy. However, this does not effect our final qualitative results.

For small $z$ the high-energy scalar masses are not-universal. At 
low-energy the spectrum is:
\begin{equation}
\tilde{m}_{g}\sim \tilde{m}_{q} > \tilde{m}_{\ell} > \tilde{m}_{\gamma}.
\end{equation}
The ratio between squark and slepton masses is about 3.5. We can easily 
see that the squark masses are quite degenerate. The slepton masses are
degenerate but to a lesser extent. This degeneracy is enough to avoid too
large FCNC even for vanishing $z$: $\tilde{m}_{3/2} \rightarrow 0$
$(K''\rightarrow 0)$ (here we compare to limits given in  
ref.~\cite{a:ggms96,a:m99} for $\tilde{m}_{q}\simeq 1\; TeV$). However,
the Higgs sector parameters require that $\tilde{m}_{3/2}$ is of order of
the electro-weak (EW) scale. In the above case this requires $z\sim 1$
(eqs.~(\ref{eq:m2z})). 

A heavy gaugino scenario with $X^{o}_{q}=4$ and $z\sim 1$ can be realized
for example with $\tilde{m}\sim \tilde{m}_{3/2}\sim 100\; GeV$, $K''\sim
\lambda^{2}-\lambda^{3}$ and $-\frac{\left< 
D_{X}\right>}{|\left<F_{s}\right>|^{2}}\sim\lambda^{3}$. 

\subsection{Light Gauginos: $\tilde{m}_{1/2}^{o} \ll
\tilde{m}^{o}_{q,\ell}$}

Light gaugino scenarios are limited by the following conditions:
\begin{eqnarray}
{\rm Naturalness}:& \hspace{1cm} &\tilde{m}_{t}\lsim 1\; TeV,\\
{\rm Experiment}:& &\tilde{m}_{g}\gsim 0.2\; TeV.
\end{eqnarray}
This imposes:
\begin{equation}
X_{q}>0.04\hspace{0.5cm} \rightarrow \hspace{0.5cm} X^{o}_{q} > 0.0045.
\label{eq:xqphlim}
\end{equation}
(The scenario in  ref.~\cite{a:bccm98} violates the above bound.)\\
The gauginos are light when the following relation is fulfilled
(eq.~(\ref{eq:msq})):
\begin{equation}
\frac{\left<F_{S}\right>^{2}}{\left<S+S^{*}\right>^{2}}\ll
\tilde{m}^{2}_{3/2}-q_{i}\left< D_{X}\right>.
\end{equation}
The mass-squared matrices of the sfermions are similar to the ones given 
in eqs.~(\ref{eq:hesq1})-(\ref{eq:hesl}), because the corrections coming 
from the gaugino masses are very small. We see that the squark and slepton 
masses-squared are of the same size.

We take for example the case $X^{o}_{q}=0.04$, with $X_{q}=0.3$ at
low-energy. In order to avoid too large FCNC we need $z>30$ (here again we
compare to limits given in ref.~\cite{a:ggms96} for $\tilde{m}_{q}\simeq
1\;TeV$; lighter squarks would have required larger $z$ in order to avoid 
too large contributions to FCNC). The slpetons are heavy enough and $z$
is large enough so that, with our choice of charge assignments, there is
no stronger limit on $z$ coming from the lepton sector. This is a scenario
of universality:
\begin{equation}
\tilde{m}^{2}_{3/2}\gg -\left<D_{X}\right> \hspace{1cm} {\rm and}
\hspace{0.5cm} K'' \gg O(\lambda).
\end{equation}
The typical spectrum is:
\begin{equation}
\tilde{m}_{q}\sim \tilde{m}_{\ell} > \tilde{m}_{g} > \tilde{m}_{\gamma}.
\end{equation}
Taking $X^{o}_{q}=0.04$ and $z\sim 31$ we can have for example 
$\tilde{m}\sim 170\; GeV$, $\tilde{m}_{3/2}\sim 900\;GeV$, $K''\sim
\lambda^{-1}$ and $-\frac{\left< 
D_{X}\right>}{|\left<F_{s}\right>|^{2}}\sim \lambda^{2}$. 

\subsection{General Analysis}

\begin{figure}[hbct]
\begin{center}
\mbox{\epsfig{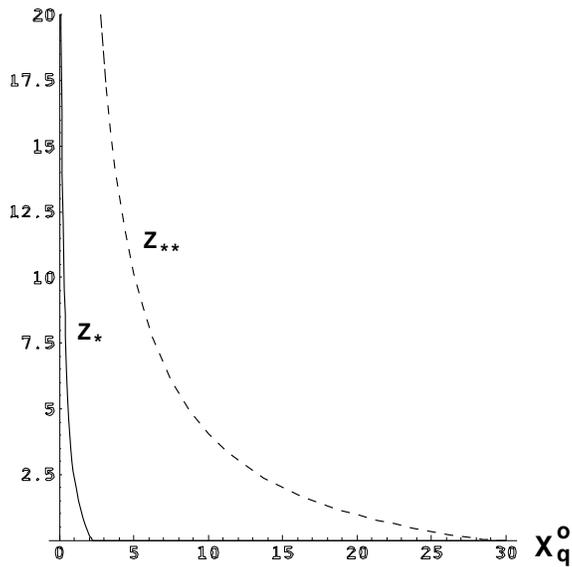}}
\end{center}
\caption{The minimal value of $z$ required, as a function of $X_{q}^{o}$, 
in order to produce a viable low-energy spectrum, calculated from the
contributions to the real $(z_{*})$ and imaginary $(z_{**})$ parts of 
$(\delta_{12}^{d})_{LL}(\delta^{d}_{12})_{RR}$ (for
$\tilde{m}_{q}=1\;TeV$).}
\label{f:zsXq}
\end{figure}

Once $X_{q}^{o}$ is chosen ($X_{q}^{o}>0.0045$), we can find a minimal
value of $z$, $z_{*}$, such that for $z>z_{*}$ there are models in which
$U(1)_{X}$ is the only horizontal symmetry, and there is no problem with
too large FCNC. Requiring that the degeneracy between the low-energy
squarks solves the supersymmetric $\epsilon_{K}$ problem~\cite{a:gnr97}
(allowing CP violating phases of $O(1)$), we find a stronger lower bound
on $z$ which we denote by $z_{**}$. $z_{*}$ and $z_{**}$ are shown in
fig.~\ref{f:zsXq} as a function of $X_{q}^{o}$ (the calculation is for
$\tilde{m}_{q}= 1\;TeV$; if the sfermions are lighter, the lower
limit on $z$ is higher). As can be seen in the graph, the heavier the
gauginos are (the larger $X_{q}^{o}$ is), the smaller the universal 
contribution to scalar masses ($z$) is allowed to be. 

\begin{figure}[hbct]
\begin{center}
\mbox{\epsfig{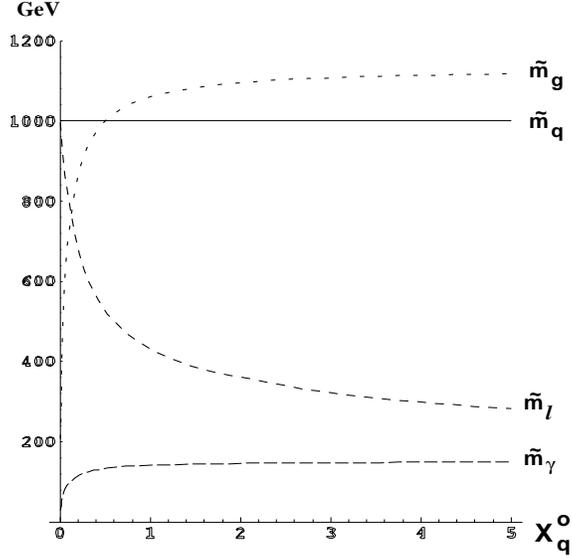}}
\end{center}
\caption{ The low-energy sparticle spectrum (in $GeV$),
$\{\tilde{m}_{g},\tilde{m}_{q},\tilde{m}_{l},\tilde{m}_{\gamma}\}=
f(X_{q}^{o})$, calculated with $z=z_{*}$. (Here $\tilde{m}_{q}=1\;TeV$ is
imposed.)}
\label{f:massXq}
\end{figure}

Taking for each $X_{q}^{o}$ the value of $z$ such that $z=z_{*}$, we get
the mass spectrum displayed in fig.~\ref{f:massXq}. 

As mentioned before, the Higgs sector parameters require $\tilde{m}_{3/2}$
to be of order of the EW scale. This sets an additional lower limit on 
viable values of $z$. Taking $\tilde{m}_{q}= 1\; TeV$ we find for 
$X^{o}_{q}\sim 2$ the bound $z\geq O(\lambda)$, rising for $X^{o}_{q}\sim
4$ to $z\geq O(1)$.

An additional limit should be set, on the value of the ratio 
$\tilde{m}^{o}_{1/2}/\tilde{m}_{q}^{o}$, in order to avoid problems with
a potential that is not bounded from below~\cite{a:clm96,a:bbc96}. This is
roughly $X^{o}_{q}\lsim 4$. One should notice that this bound implies that
scenarios in which the degeneracy between the low energy squarks solves
the SUSY $\epsilon_{K}$ problem require $z>12$ (see $z_{**}$ in
fig.~\ref{f:zsXq}).

The ratio of the vevs is given by (see
eqs.~(\ref{eq:mo12}),(\ref{eq:Xo}),(\ref{eq:m2z})):
\begin{equation}
-\frac{\left< D_{X}\right>}{|\left< F_{S}\right> |^{2}}= 
\frac{1}{\frac{\tilde{m}_{q}^{o\:{2}}}{\tilde{m}^{2}}X_{q}^{o}
\left<S+S^{*}\right>^{2}}.\label{eq:ratDF}
\end{equation}
Choosing $X_{q}^{o}\;(X_{q}^{o}>O(\lambda^{2}))$ and $z\;(>z_{*})$, the
value of this ratio is found to be:
\begin{equation}
-\frac{\left< D_{X}\right>}{|\left< F_{S}\right> |^{2}} \leq
O(\lambda^{2}) \label{eq:ratDFv}
\end{equation}
(for smaller $X_{q}^{o}$ this ratio can rise up to $O(\lambda)$).

Even if additional vevs other than those appearing in eq.~(\ref{eq:vevs})
contribute to the soft terms, in particular to $\tilde{m}_{3/2}$,
fig.~\ref{f:zsXq} gives an estimate of the amount of sfermion degeneracy
required at the high-energy scale, relative to the gaugino mass, in order
to arrive at a viable scenario in which the anomalous $U(1)_{X}$ flavor
symmetry is involved in SUSY breaking, and fig.~\ref{f:massXq} gives the
resulting spectra. Eqs.~(\ref{eq:ratDF}),(\ref{eq:ratDFv}) also hold for
any composition of $\tilde{m}_{3/2}$.

Returning to the specific assumption of our model (dominance of the vevs
in eq.~(\ref{eq:vevs})), we find that, for given $X^{o}_{q}$ and $z$, the
K\"ahler potential obeys (see 
eqs.~(\ref{eq:m32}),(\ref{eq:mo12}),(\ref{eq:Xo}),(\ref{eq:m2z})):
\begin{eqnarray}
K''&=&\frac{f(X_{q}^{o},z)}{\left<S+S^{*}\right>^{2}} \label{eq:K''a}\\
{\rm with}\hspace{2cm}
f(X_{q}^{o},z)&=&\frac{3z}{\frac{\tilde{m}_{q}^{o\:{2}}}{\tilde{m}^{2}}X_{q}^{o}}
<\frac{3}{X_{q}^{o}}.
\nonumber
\end{eqnarray}
The bound in eq.~(\ref{eq:xqphlim}) implies $K''\leq  O(\lambda^{-2})$. 

\subsection{A Viable Scenario}

The above analysis is directed at building phenomenologically viable
models without imposing any relations between the different vevs. In any
case there is a large degeneracy, but the minimal required degeneracy to
allow for phenomenologically viable models is given by parameters on the
$z=z_{*}$ line. There are possible charge assignments for the leptons,
including the one we chose, that predict (for particular $z$ and
$X_{q}^{o}$) $Br(\mu\rightarrow e\gamma)$ at the experimental limit.

In all of the above the anomalous $U(1)_{X}$ symmetry plays an important role
in explaining the flavor parameters of the fermion sector. We are
interested, however, in models in which this $U(1)_{X}$ plays an
important role also in SUSY breaking. In particular, we assume that 
$\left< D_{X}\right>$ is the dominant source of non-universality in the
scalar masses and we are interested in the case that the contribution is
not negligibly small compared to the universal one, say, $z<10$. This
requires, according to fig.~\ref{f:zsXq}, heavy gauginos 
($X_{q}^{o}\gsim0.3$). This last statement is correct also when there are
additional contributions to $\tilde{m}_{3/2}$. Under the specific
assumption made here, that is to say dominance of the vevs in
eq.~(\ref{eq:vevs}), this implies $K''\leq O(\lambda)$ (see
eq.~(\ref{eq:K''a})) and consequently $K''\lsim (-K')$.

A viable scenario with $X_{q}^{o}=2.5$ and $z=0.4$ can be realized for 
example with $\tilde{m}_{q}\sim 1\;TeV$, $\tilde{m}\sim 145\;GeV$, 
$\tilde{m}_{3/2}\sim 90\;GeV$, $\tilde{m}_{l}\sim 330\;GeV$, $K''\sim
\lambda^{3}$ and $-\frac{\left< D_{X}\right>}{|\left<
F_{s}\right>|^{2}}\sim \lambda^{3}$.
 
The scenario requires the following relations to be obeyed:
\begin{itemize}
\item $-\frac{\left< D_{X}\right>}{|\left<F_{s}\right>|^{2}}=
 O(\lambda^{2})- O(\lambda^{3})$,
\item $-\delta_{GS}K'= O(\lambda^{2})$, with $|K'|= O(\lambda)$ and
 $\delta_{GS}=O(\lambda)$,
\item $K''=\frac{f(X^{o}_{q},z)}{\left< S+S^{*}\right>^{2}}$ as given in
 eq.~(\ref{eq:K''a}), leading to $K''\leq O(\lambda)$.
\end{itemize}
Is there a K\"ahler potential that can fulfill these requirements? This
question is beyond the scope of this work. However, it does seem to be a
good sign that even the weak coupling form of the K\"ahler potential,
$K=-\ln(S+S^{*})$, comes close to fulfilling these requirements.

\section{Conclusions \label{s:conclu}}

In this work a bottom-up approach is used in order to characterize the
different possible models with an anomalous $U(1)_{X}$ flavor symmetry
that is involved in SUSY breaking. Assuming that the leading contributions
to the different soft-terms in the model come from the following vevs:
\begin{equation}
\left<\phi_{-}\right>, \hspace{1cm} \left<F_{S}\right>, \hspace{1cm}
\left<D_{X}\right>,
\end{equation}
the characteristics of the low-energy spectrum are given. The $U(1)_{X}$
horizontal symmetry explains naturally the smallness and hierarchy of the 
observed flavor parameters of fermions. Only mild alignment is produced.
This leads, through the demand for suppression of FCNC, to scenarios with
degenerate sfermions at low-energy. The horizontal symmetry is not 
manifest in the sfermion spectrum. The different possibilities for
producing such a degenerate spectrum were described, ranging from
universality scenarios to scenarios with heavy gaugino. The different
scenarios result in different relative sizes of squarks, sleptons and
gaugino masses. This class of models allows for $Br(\mu\rightarrow
e\gamma)$ at the current experimental limit. For all the 
phenomenologically viable scenarios ($X_{q}^{o}>O(\lambda^{2})$), the
following is required:
\begin{equation}
-\frac{\left< D_{X}\right>}{|\left< F_{S}\right> |^{2}}\leq 
O(\lambda^{2}).
\end{equation}
It seems impossible to build models in which the degeneracy is strong
enough to solve the supersymmetric $\epsilon_{K}$ problem while having the
relevant CP violating phases of $O(1)$. Approximate CP~\cite{a:en98} is a 
possible solution to this problem.

Although there are different phenomenologically viable scenarios, the 
contributions of $\left< D_{X}\right>$ to scalar masses at high energy are
significant only with heavy gauginos. In that case the dilaton K\"ahler
potential needs to have the
following features: 
\begin{equation}
|K'|=O(\lambda),\hspace{2cm} K''\leq O(\lambda).
\end{equation}

We were led to the above scenarios without assuming any a-priori 
relationship between the different vevs. Whether it is possible to build a
complete model in which the leading contributions come only from these
three vevs, and with the required relations between them, is beyond the
scope of this work. 

While this work was near completion, a paper appeared~\cite{a:km99} that 
deals with models similar to those described above. There, the starting
point is an analysis of the K\"ahler potential, which is an approach
different from ours. The implications of such models on $Br(\mu\rightarrow
e\gamma)$ were stressed.\\

{\bf Acknowledgments}

I am grateful to Yossi Nir and Yael Shadmi for enlightening and helpful 
discussions and for comments on the manuscript. I also thank Jan Louis for
a useful conversation.

{}

\begin{thebibliography}{99}
\bibitem{a:dp96} G. Dvali and A. Pomarol, Phys. Rev. Lett. 77 (1996) 3728,
 hep-ph/9607383.
\bibitem{a:bd96} P. Binetruy and E. Dudas, Phys. Lett. B389 (1996) 503,   
 hep-th/9607172.
\bibitem{a:adm98} N. Arkani-Hamed, M. Dine and S.P. Martin, Phys. Lett.
 B431 (1998) 329, hep-ph/9803432.
\bibitem{a:gs84} M. Green and J. Schwarz, Phys. Lett. B149 (1984) 117.
\bibitem{a:ir94} L.E. Ibanez and G.G. Ross, Phys. Lett. B332 (1994) 100, 
 hep-ph/9403338.
\bibitem{a:br95} P. Binetruy and P. Ramond, Phys. Lett. B350 (1995) 49,   
 hep-ph/9412385.
\bibitem{a:nw97} A.E. Nelson and D. Wright, Phys. Rev. D56 (1997) 1598,
 hep-ph/9702359.
\bibitem{a:ns93} Y. Nir and N. Seiberg, Phys. Lett. B309 (1993) 337,
 hep-ph/9304307.
\bibitem{a:dg95} S. Dimopoulos and G.F. Giudice, Phys. Lett. B357 (1995)
 573, hep-ph/9507282.
\bibitem{a:ckn96} A.G. Cohen, D.B. Kaplan and A.E. Nelson, Phys. Lett.
 B388 (1996) 588, hep-ph/9607394.
\bibitem{a:am97} N. Arkani-Hamed and H. Murayama, Phys. Rev. D56 (1997)  
 6733, hep-ph/9703259.
\bibitem{a:hkn98} J. Hisano, K. Kurosawa and Y. Nomura, Phys. Lett. B445
 (1999) 316, hep-ph/9810411.
\bibitem{a:sw83} S.K. Soni and H.A. Weldon, Phys. Lett. B126 (1983) 215.
\bibitem{a:gm88} G.F. Giudice and A. Masiero, Phys. Lett. B206 (1988) 480.
\bibitem{a:kl93} V. S. Kaplunovsky and J. Louis, Phys. Lett. B306 (1993) 
 269, hep-th/9303040.
\bibitem{a:lns94} M. Leurer, Y. Nir and N. Seiberg, Nucl. Phys. B420
 (1994) 468, hep-ph/9310320.
\bibitem{a:n95} Y. Nir, Phys. Lett. B354 (1995) 107, hep-ph/9504312.
\bibitem{a:ceklp95} D. Choudhury, F. Eberlein, A. K\"onig, J. Louis and S.
 Pokorski, Phys. Lett. B342 (1995) 180, hep-ph/9408275.
\bibitem{a:ggms96} F. Gabbiani, E. Gabrielli, A. Masiero and L.
 Silvestrini, Nucl. Phys. B477 (1996) 321, hep-ph/9604387. 
\bibitem{a:m99} MEGA Collaboration, hep-ex/9905013.
\bibitem{a:bccm98} T. Barreiro, B. de Carlos, J.A. Casas and J.M. Moreno,
 Phys. Lett. B445 (1998) 82, hep-ph/9808244. 
\bibitem{a:gnr97} Y. Grossman, Y. Nir and R. Rattazzi, in {\it Heavy
 Flavours II}, eds. A.J. Buras and M. Lindner, World Scientific Publishing
 Co., hep-ph/9701231. 
\bibitem{a:clm96} J.A. Casas, A. Lleyda and C. Munoz, Nucl. Phys. B471
 (1996) 3, hep-ph/9507294.
\bibitem{a:bbc96} H. Baer, M. Brhlik and D. Castano, Phys. Rev. D54 (1996)
 6944, hep-ph/9607465.
\bibitem{a:en98} G. Eyal and Y. Nir, Nucl. Phys. B528 (1998) 21,
 hep-ph/9801411.
\bibitem{a:km99} K. Kurosawa and N. Maekawa, hep-ph/9902469.
\end{thebibliography}
\end{document}